\documentclass[aps,prb,twocolumn,superscriptaddress]{revtex4-1}

\usepackage{amssymb}
\usepackage{suffix}
\usepackage{float}
\usepackage{mathtools}
\usepackage[utf8]{inputenc}
\usepackage{booktabs}
\usepackage{tabularx}
\usepackage{booktabs}
\usepackage{cases}
\usepackage[multiple]{footmisc}
\usepackage{dcolumn}
\usepackage{physics}
\usepackage{color,soul}
\usepackage{rotating}
\usepackage{perpage}
\usepackage{xcolor}
\usepackage{soul}
\usepackage{tikz}
\usepackage[T1]{fontenc}
\usepackage{etoolbox}
\usepackage{graphics}
\usepackage{siunitx}
\usepackage[hyperindex,breaklinks]{hyperref}
\usepackage{float}	
\usepackage{collref}
\usepackage{multirow}
\usepackage{mathtools}
\usepackage{bm}
\usepackage{url}

\usepackage{tikz}
\usepackage{tikz-3dplot}
\usepackage{changes}

\newcommand{\blue}[1]{\textcolor{blue}{#1}}

\newcommand{\quotes}[1]{``#1''}

\begin{document} 
	\title{Truncated Wigner approximation for the bosonic model of large spin baths}
	\author{Mohsen Yarmohammadi}
	\email{mohsen.yarmohammadi@utdallas.edu}
	\affiliation{Department of Physics, The University of Texas at Dallas, Richardson, Texas 75080, USA}
	\affiliation{Condensed Matter Theory, TU Dortmund University, Otto-Hahn Straße 4, 44221 Dortmund, Germany}
	\author{Katrin Bolsmann}
	\email{katrin.bolsmann@tu-dortmund.de}
	\affiliation{Condensed Matter Theory, TU Dortmund University, Otto-Hahn Straße 4, 44221 Dortmund, Germany}
	\author{Yvonne Ribbeheger}
	\email{yvonne.ribbeheger@tu-dortmund.de}
	\affiliation{Condensed Matter Theory, TU Dortmund University, Otto-Hahn Straße 4, 44221 Dortmund, Germany}
	\author{Timo Gr\"a\ss er}
	\email{timo.graesser@tu-dortmund.de}
	\affiliation{Condensed Matter Theory, TU Dortmund University, Otto-Hahn Straße 4, 44221 Dortmund, Germany}
	\author{Götz S. Uhrig}
	\email{goetz.uhrig@tu-dortmund.de}
	\affiliation{Condensed Matter Theory, TU Dortmund University, Otto-Hahn Straße 4, 44221 Dortmund, Germany}
	
	\date{\today}
	\begin{abstract}
The central spin model has a wide applicability, it is ideally suited to describe a small quantum system, for instance a quantum bit, in contact to a bath of spins, e.g., nuclear spins, or other small quantum systems in general. According to previous work~[R\"ohrig \textit{et al.}, Phys. Rev. B {\bf 97}, 165431 (2018)], a large bath of quantum spins can be described as a bath of quantum harmonic oscillators. But the resulting quantum model is still far from being straightforward solvable. Hence we consider a chain representation for the bosonic degrees of freedom to study how well a truncated Wigner approximation of the effective model of harmonic oscillators works in comparison with other approximate and exact methods. Numerically, we examine the effect of the number of bath spins and of the truncation level, i.e., the chain length.
	\end{abstract}
	\maketitle
	{\allowdisplaybreaks
		\section{Introduction}
		\label{s:intro}

The central spin model~(CSM) is a well-known model describing the interaction of a single \quotes{central} spin with surrounding spins~\cite{refId0,gaudin1983fonction}, for instance, the interaction of the spin of a localized electron with nuclear spins in quantum dots~\cite{PhysRevA.57.120,PhysRevLett.100.236802,PhysRevB.70.195340}. In view of 
the intense search for physical realizations of quantum bits~\cite{nielsen2010quantum}, a localized electron in a quantum dot can be seen as a two-level system and thus as a promising candidate for quantum bits~\cite{Greilich2009,PhysRevB.74.205415,PhysRevLett.102.167403}. The CSM
is a quantum many-body system and major progress has been made to understand 
its properties in its applications for phenomena
 in material science and quantum information technology~\cite{PhysRevX.5.011022,Smith2019,doi:10.1080/00018732.2010.505452,PhysRevB.74.035322}. Polarization recovery in a longitudinal field~\cite{PhysRevLett.94.116601,PhysRevB.102.235413}, nuclei-induced frequency focusing~\cite{PhysRevB.98.024305,doi:10.1126/science.1146850,PhysRevB.98.155318}, spin precession mode locking~\cite{PhysRevB.102.115301,PhysRevB.85.125304}, the effect of spin inertia~\cite{PhysRevB.98.125306,PhysRevResearch.1.033189}, spin noise~\cite{PhysRevB.86.115308,PhysRevB.98.045307,PhysRevB.97.195311,PhysRevB.93.205429,PhysRevB.89.045317}, and many other effects belong to the particularly rich physics of the CSM. Furthermore, the CSM is also used to understand the dynamics of quantum sensors~\cite{doi:10.1063/1.5010282} 
which helps to reach high sensitivities.

For a finite, not too large number of bath spins~\cite{https://doi.org/10.1002/pssb.200945229}, it is possible to use the Bethe ansatz~\cite{refId0,farib13a,farib13b} to diagonalize the CSM Hamiltonian and to analyze
rigorous restrictions of the central spin dynamics stemming from conserved quantities~\cite{uhrig14a,PhysRevB.94.094308}.
If all couplings are equal the CSM reduces to the so called \textit{box model} allowing one to compute the spin dynamics for large spin baths essentially analytically~\cite{PhysRevB.70.014435,Kozlov2007,Bortz_2007}. However, the complexity of the CSM in practical applications is related mainly to the electron spin decoherence when interacting with an (almost) infinite number of nuclei spins~\cite{Yang_2016,PhysRevB.65.205309,PhysRevLett.88.186802,PhysRevB.70.195340,PhysRevB.74.195301,PhysRevLett.109.140403,PhysRevLett.98.077601}. In this scenario, the initial polarization and information on the spin state is quickly and irreversibly lost. 

To describe this decoherence of the central spin and to conceive 
strategies against it, various approaches have been conceived. Density-matrix renormalization group~(DMRG) deals with up to 1000 spins, but only up to relatively short times~\cite{PhysRevB.88.155305,PhysRevB.90.064301} due to the fast growth of entanglement. The linked-cluster and cluster-correlation expansions~\cite{PhysRevLett.93.120503,PhysRevB.75.125314,PhysRevB.78.085315,PhysRevLett.120.220604} investigate the long-time spin decoherence, but of finite, relatively small spin baths. Moreover, considering the nuclear-electric quadrupolar interactions for a few spins, the spin-noise spectrum at various timescales has been calculated using Chebyshev polynomials~\cite{PhysRevLett.115.207401,PhysRevB.89.045317,Smirnov_2021}. Furthermore, a coherent interface between electron and nuclear spins was recently developed~\cite{doi:10.1126/science.aaw2906} with the vision to realize long-lived quantum memory. 
		
Although a classical description of CSM with a large-enough number of nuclear spins can be justifed over a long time, it neglects all quantum mechanical aspects~\cite{PhysRevB.88.155305,PhysRevB.90.064301} which are vital for quantum bits. This originates from the fact that the central spin is a truly quantum mechanical object and its back-action on the bath spins is not classical. The truncated Wigner	approximation~(TWA)~\cite{POLKOVNIKOV20101790} is a general semi-classical approach  in which quantum fluctuations 
are partly taken into account through random initial conditions for the classical equations of motion.
 Although the equations of motion themselves are still purely classical, correlations and the probabilities of quantum
measurements can be simulated to a certain degree. The TWA has often been used to 
simulate the dynamics of the CSM~\cite{PhysRevLett.114.045701,PhysRevX.5.011022,PhysRevB.96.054415}. The spins are taken as classical 
vectors precessing around local classical fields.
We abbreviate this semi-classical approach to spins sTWA. It can be implemented for moderate
numbers of spins ($N\approx 200$) if one has to simulate long times. Experimentally, the bath sizes
range from $10^4$ to $10^6$ still exceeding numerical resources by far even though
a hierarchical chain representation based on generalized Overhauser fields helps to reconcile
large spin baths and long-time simulations~\cite{PhysRevB.96.054415}.


In this framework, a fully quantum mechanical approach~\cite{PhysRevB.97.165431} based on iterated equations of 
motion~(iEoM) has been suggested for large spin baths. The asset of this approach is that it is particularly suited
to capture very large or even infinitely large spin baths. The bath of spins is mapped
to a bath of hierarchically coupled bosons and the central spin is mapped to 
a four-dimensional impurity. But the fully quantum mechanical evaluation of the dynamics of the 
effective bosonic model for long times represents still a tremendous challenge. Hence, it is interesting to 
study approximate ways to treat this effective bosonic model. 

In this work, we study the application of the TWA to the mapped effective bosonic model resulting from iEoM~\cite{PhysRevB.97.165431}, i.e., to the harmonic 
oscillators. The impurity is described by two spins with $S=1/2$ which, in turn, are treated as classical vectors.
In order to distinguish this TWA from the one resulting from the classical treatment
of the spins we call it bosonic TWA~(bTWA). Clearly, the bTWA would 
remove the restrictions on the maximum number of bosonic modes which can be simulated. 
The immediate aim is to describe the experimental spin noise spectra~\cite{PhysRevB.93.205429,PhysRevLett.110.176601,PhysRevLett.104.036601}.
To benchmark the bTWA, we compare our data to data from some of the 
above-mentioned techniques under the same conditions.
		
This paper is organized as follows. In Sec.~\ref{s:initial_model}, we review the CSM and  in 
Sec.~\ref{s:eff_model}, we present its bosonic formulation. In Sec.~\ref{s:results}, 
we present our results and compare them with results from
other techniques. Finally, the paper is summarized in Sec.~\ref{s:summary}.

		\section{Initial Model}
		\label{s:initial_model}
		
		In this section, we briefly introduce the CSM. For our proof-of-principle study, we restrict ourselves to the paradigmatic isotropic version of the CSM. This implies that we neglect dipole-dipole interaction~\cite{PhysRevB.65.205309,Schliemann_2003}, quadrupolar couplings~\cite{bulut12,bulut14,Chekhovich2012,PhysRevLett.109.166605,PhysRevLett.115.207401}, and spin-orbit couplings~\cite{doi:10.1126/science.1148092,PhysRevB.90.245305,PhysRevB.61.12639,PhysRevB.64.125316,PhysRevLett.93.016601} of the nuclear spins which usually become relevant on very long timescales. We start with the CSM comprising a central spin $\hat{\vec{S}}_{0}$ with $S=1/2$ interacting through the hyperfine coupling with a bath of $N$ spins $\hat{\vec{S}}_{i}$. The Hamiltonian reads
		\begin{equation}		\label{eq_1}
		\hat{\mathcal{H}} = {} \sum^{N}_{i=1} \,J_i \,\hat{\vec{S}}_0 \cdot \hat{\vec{S}}_i\, ,
		\end{equation}
		where $J_{i}$ denotes the hyperfine coupling of the $i$-th spin in the bath. 
		In electronic quantum dots, the coupling constants $J_{i}$ are proportional to the probability  that the 
		electron is present at the site of the nucleus $i$~\cite{PhysRevB.65.205309,Schliemann_2003} which is given by the modulus squared of the electronic wave function. It is convenient to define a composite field for the effect of the bath spins, 
		$\hat{\vec{B}} = {} \sum^{N}_{i=1}\, J_i\,\hat{\vec{S}}_{i}$, which is called the Overhauser field. With its
		help, the Hamiltonian can simply be rewritten as $\hat{\mathcal{H}}=\hat{\vec{S}}_{0}\cdot\hat{\vec{B}}$.
		
		Let us consider an infinite spin bath~($N\to\infty$) with decreasing couplings.
		We consider the generic parametrization $J_i = C \exp(-i \gamma)$~\cite{farib13a,farib13b,PhysRevB.88.085323,PhysRevB.94.094308,PhysRevB.98.024305,PhysRevB.96.054415} with $i \in [1,N]$, where the prefactor $C$ sets the energy scale. 
		For $\gamma>0$, the exponential term is decreasing with $i$. The meaning of $\gamma$ is elucidated by the
		following argument. Even if $N\to\infty$, there is only a finite number of bath spins which 
		is  appreciably coupled to the central spin. We denote this number by $N_{\text{eff}}$ and define
		it by the ratio of the squared sum of all couplings and the sum of all squared couplings~\cite{PhysRevB.65.205309,Schliemann_2003,PhysRevB.88.155305,PhysRevB.90.064301,PhysRevB.94.094416,PhysRevB.96.054415}, i.e., $N_{\rm eff} := {} \Big(\sum^{N}_{i=1}J_{i}\Big)^2/J^2_{\mathrm Q}$,
		where $J^2_{\mathrm Q} := \sum^{N}_{i=1}J^{2}_{i}$. Inserting our parametrization $J_i$ into $N_{\rm eff}$ in the limit $N\to\infty$, we find for small values $\gamma$
		\begin{equation}	
		\label{eq_2}
		\begin{aligned}
		N_{\text{eff}}=\frac{2}{\gamma}+\mathcal{O}(\gamma)\,.
		\end{aligned}
		\end{equation}
		So $\gamma$ is about twice the inverse number of effectively coupled spins.
		The electron spin in quantum dots is coupled to a very large number of bath spins~\cite{PhysRevB.65.205309,Schliemann_2003,doi:10.1063/1.1850605,PhysRevB.78.045315}, 
		$N_{\text{eff}}\approx10^{4}$ to $10^{6}$, so, $\gamma\approx10^{-4}$ to $10^{-6}$ is a realistic estimate. Moreover, we set the energy scale for all simulations by requiring $J_{\mathrm{Q}}=1$. This results in $C\simeq\sqrt{2\gamma}\approx10^{-2}$ to $10^{-3}$, which is a very small number implying that the contribution of an individual bath spin is negligible.
		Only suitable sums over all spins have a sizable impact. 
		In contrast, for large $\gamma$, we deal with a small number of bath spins, see Eq.~\eqref{eq_2}, and 
		the dynamics of the central spin can be determined using 
		fully quantum mechanical descriptions~\cite{PhysRevLett.115.207401,PhysRevB.89.045317,PhysRevLett.120.220604}.


		\section{Effective model and semi-classical approach}\label{s:eff_model}
		
		In what follows, we sketch the mapping of the spin bath on a bosonic bath~(iEoM~\cite{PhysRevB.97.165431}). Then, we introduce the semi-classical approach bTWA based on a hierarchical chain representation to describe 
		the \textit{long-time} spin dynamics. 

		\subsection{Objective}\label{s:objective}
		
		We begin with the application of the Heisenberg equation of motion to the CSM, $\partial_{t} \hat{\mathcal{A}}={\tt i}[\hat{\mathcal{H}},\hat{\mathcal{A}}]$~(throughout the present work, $\hbar$ is set to unity), where $\hat{\mathcal{A}}$ are operators of the CSM forming a suitable operator basis for the products of all components of spin operators at all sites~\cite{PhysRevB.97.165431}. In the end, we are interested in the $\alpha$ component of the spin-spin autocorrelation function of the central spin at infinite temperature
		\begin{equation}		\label{eq_3}
		S^\alpha (t)={}\langle \hat{S}^\alpha_{0}(t)\hat{S}^\alpha_{0}(0)\rangle\, ,
		\end{equation} 
		for small values of the parameter $\gamma$ corresponding to large spin baths. In particular, the long-term behavior of $S^z(t)$ provides information about the fate of state with the central spin aligned along the $z$-axis initially, i.e., at $t=0$. Assuming infinite temperature is well justified because the thermal energy in the bath 
		at temperatures of a few Kelvin is at least one order of magnitude larger than the
		individual couplings in a quantum dot~\cite{RevModPhys.85.79}. 
		
	For motivation, we provide the autocorrelation if a constant external or internal magnetic field is applied~\cite{PhysRevB.65.205309,PhysRevB.88.155305}
		\begin{equation}	\label{eq_4}
		\begin{aligned}
		\hat{\vec{S}}_{0}(t)={} &{\vec{n}}\big[{\vec{n}}\cdot\hat{\vec{S}}_{0}(0)\big]+\big\{\hat{\vec{S}}_{0}(0)-{\vec{n}}
		\big[{\vec{n}}\cdot\hat{\vec{S}}_{0}(0)\big]\big\}\cos(Bt)\\&-\big[{\vec{n}}\times\hat{\vec{S}}_{0}(0)\big]\sin(Bt)\, ,
		\end{aligned}
		\end{equation}
		where ${\vec{n}}$ points in the direction of the magnetic field. This formula is identical to the classical one
		since ${\vec{B}}$ is a classical vector and the equations of motion are linear in the spin operators. 
		Obviously,  powers of $B$ up to infinite order occur so 
		that a suitable operator basis needs operators including high powers of the Overhauser field if we
		want to capture its intrinsic quantum character and the ensuing dynamics. 
			\begin{figure*}[t]
			\centering
			\includegraphics[width=0.7\linewidth]{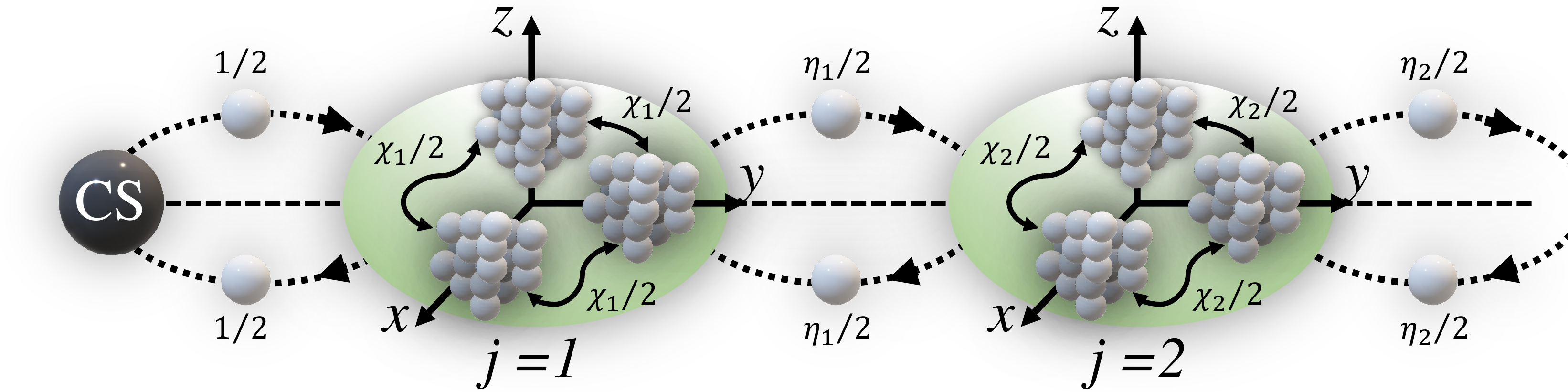}
			\caption{Sketch of the CSM described by Eqs.~\eqref{eq_6a} and~\eqref{eq_6b}. The central spin and the 
				bosons in the chain are shown by the black and light gray solid spheres, respectively. The solid two-sided arrows 
				inside the boxes illustrate the couplings $\chi_{j}/2$ between bosons of different flavors at the same
				site of the chain, while the dotted ones indicate the couplings $\eta_{j}/2$ between bosons on adjacent sites.}
			\label{fig1} 
		\end{figure*}		
	
		If one neglects the dynamics
		of the Overhauser field completely, the frozen Overhauser field approximation is retrieved for which 
		one averages over all random directions and random strengths of the 
		Overhauser field~\cite{PhysRevB.65.205309,PhysRevB.88.155305} yielding
		\begin{equation}	
		\label{eq:frozen_overhauser}
		S^\alpha(t) = \frac{1}{12}\left[2e^{-J_{\mathrm Q}^2t^2/8}(1-J_{\mathrm Q}^2t^2/4)+1\right]\, .
		\end{equation} 
		This analytic result is convenient as reference, see the figures below.

		
		\subsection{Effective model with higher powers of the Overhauser field
		\label{s3b}}
		
		The orthogonal Hermite polynomials of the Overhauser field and similar composite weighted sums
		of the bath spin have been introduced by R\"ohrig \textit{et al.}~\cite{PhysRevB.97.165431} as suitable operator basis. These polynomials are orthogonal for a Gaussian weight function~\cite{abramowitz1965handbook}, and can be applied to different components of generalized Overhauser fields by the recursive relation
		\begin{equation}	\label{eq_5}
		\begin{aligned}
		G^{\alpha}_{j}H_{n}(G^{\alpha}_{j})=\sqrt{n}H_{n-1}(G^{\alpha}_{j})+\sqrt{n+1}H_{n+1}(G^{\alpha}_{j}),
		\end{aligned}
		\end{equation}
		where $\alpha=\{x,y,z\}$ and $H_{0}(G^{\alpha}_{j})=1$ by definition. The polynomials
		$H_{n}(G^{\alpha}_{j})$ are the Hermite polynomials of degree $n$ in the generalized Overhauser field vectors 
		$\vec{G}_{j}$. These fields are defined	by
		\begin{equation}
		G^{\alpha}_{j} := 2\sum_{i=1}^N \mathcal{P}_{j}(J_{i}) \hat S^\alpha_i\, ,
		\end{equation}
		where the real orthogonal polynomials $\mathcal{P}_{j}(x)$ are defined such that they comply with the orthogonality relation~\cite{PhysRevB.97.165431,PhysRevB.96.054415}
		\begin{equation}
		\label{eq:ortho_P}
		\delta_{j,m}=\sum_{i=1}^N \mathcal{P}_{j}(J_i) \mathcal{P}_{m}(J_i)\, .
		\end{equation}
 The polynomials $\mathcal{P}_{j}(J_{i})$ describe the weight of each bath spin  $\vec{S}_{i}$. The established EoM for this basis of operators tells us that a single $H_{n}(G^{\alpha}_{j})$ is transformed into the terms $\sqrt{n}H_{n-1}(G^{\alpha}_{j})$ and $\sqrt{n+1}H_{n+1}(G^{\alpha}_{j})$. This is identical to the effect of an annihilation ($\hat{a}$) and creation ($\hat{a}^{\dagger}$) bosonic operator, respectively, applied to the eigenstates $\ket n$ of an harmonic oscillator.
		
		Eventually, a quantum mechanical representation of large spin baths by means of the iEoM for the generalized Overhauser fields including an external magnetic field has been obtained and developed, see 
		Ref.~\onlinecite{PhysRevB.97.165431} for further details. It is shown that in the limit $N\to\infty$ the isotropic CSM can be mapped onto a 
		four-dimensional impurity coupled to a non-interacting bosonic bath yielding the effective Hamiltonian 
		$\hat{\mathcal{H}}_{\text{eff}} = \hat{\mathcal{H}}^{\text{CS}}_{\text{eff}}+\hat{\mathcal{H}}^{\text{ch}}_{\text{eff}} + \hat{\mathcal{H}}^{\rm Z}_{\rm eff}$ in the presence of an external Zeeman magnetic field $h$ along the $z$-direction. It is given by
		\begin{subequations}		\label{eq_6}
			\begin{align}
			\hat{\mathcal{H}}^{\text{CS}}_{\text{eff}} = 
			{} &\frac{1}{2}\sum_{\alpha=1}^{3}\hat{K}_{\alpha}\big(\hat{a}^{\dagger}_{1,\alpha}+\hat{a}_{1,\alpha}\big)\, ,
			\label{eq_6a}\\
			\hat{\mathcal{H}}^{\text{ch}}_{\text{eff}}={} &\frac{\texttt{i}}{2}\sum_{j=1}^{N_{\text{tr}}}\sum_{\alpha,\beta,\delta=1}^{3}\epsilon_{\alpha\beta\delta}\hat{M}_{\beta}\big[\chi_{j}\hat{a}^{\dagger}_{j,\delta}\hat{a}_{j,\alpha}\notag\\\qquad {}&\qquad \qquad+\eta_{j}(\hat{a}^{\dagger}_{j+1,\delta}\hat{a}_{j,\alpha}-\hat{a}^{\dagger}_{j,\alpha}\hat{a}_{j+1,\delta})\big]\, 
			,\label{eq_6b}\\
			\hat{\mathcal{H}}^{\rm Z}_{\rm eff} = {} & - h \hat{K}_{z}\, ,\label{eq_6c}
			\end{align}
		\end{subequations}
		where $\hat{\mathcal{H}}^{\text{CS}}_{\text{eff}}$ refers to the central spin located at the head of a bosonic chain, whereas $\hat{\mathcal{H}}^{\text{ch}}_{\text{eff}}$ acts on a bosonic chain with flavors $\alpha$ as depicted
		in Fig.~\ref{fig1}.  In the above equations, $\epsilon_{\alpha\beta\delta}$ is the Levi-Civita tensor.
		The couplings $\eta_j$ and $\chi_j$ result from the recursion of the orthogonal
		polynomials $\mathcal{P}_{j}$ which can be expressed in the matrix form
			\begin{equation}		\label{eq_7}
		\hat{\mathcal{T}}=
		\begin{pmatrix}
		\chi_{1} & \eta_{1} & 0 & 0 & \cdots\\
		\eta_{1} & \chi_{2} & \eta_{2} & 0 & \cdots\\
		0 & \eta_{2} & \chi_{3} & \eta_{3} & \cdots\\
		\vdots & \vdots & \ddots & \ddots & \ddots
		\end{pmatrix}\, ,
		\end{equation}
with $J_{i}\underline{\mathcal{P}}_{j}(J_{i})=\hat{\mathcal{T}}\underline{\mathcal{P}}_{j}(J_{i})$ using the 
 vector of polynomials $\underline{\mathcal{P}}_{j}(J_{i})=
\left[\mathcal{P}_{1}(J_{i}),\mathcal{P}_{2}(J_{i}),\cdots \mathcal{P}_{n}(J_{i})\right]^\top$. 
By definition, we have $\eta_0  = 0$. While the chain is half-infinite for an infinite bath, it is truncated at 
$j_{\rm max}$ in practical 	calculations~\cite{PhysRevB.96.054415,PhysRevB.97.165431} so that we also have 
	$\eta_{j_{\text{max}}}=0$. (In Ref.~\onlinecite{PhysRevB.96.054415}, the truncation level 
	was denoted by $N_{\rm tr} = j_{\text{max}}$.)
	
		The commutation and anticommutation of the operators of the central spin with $\hat{\sigma}_{\alpha}$~(Pauli matrices) in the chain are expressed by the matrices $\hat{K}_{\alpha}$ and $\hat{M}_{\alpha}$, respectively, with matrix elements $\langle\langle n|\hat{K}_{\alpha}|m\rangle\rangle={} \frac{1}{2}\langle\langle \hat{\sigma}_{n}|[\hat{\sigma}_{\alpha},\hat{\sigma}_{m}]\rangle\rangle$ and $\langle\langle n|\hat{M}_{\alpha}|m\rangle\rangle={} \frac{1}{2}\langle\langle \hat{\sigma}_{n}|\{\hat{\sigma}_{\alpha},\hat{\sigma}_{m}\}\rangle\rangle$ for $\{m,n\} \in \{x,y,z\}$. The notation $\langle\langle \dots \rangle \rangle$ is used for the scalar product of operators for which we use 
		$\langle\langle \hat A |\hat B\rangle \rangle := \langle \hat A^\dag \hat B\rangle_{T=\infty}$, i.e., 
 the expectation value at infinite temperature.	Straightforwardly, we find
		\begin{subequations}\label{eq_8}
			\begin{align}
			\hat{K}_\alpha = {} & {\tt i}\begin{pmatrix}
			0 & 0 & 0 & 0\\
			0 & 0 & \delta_{\alpha,z} & -\delta_{\alpha,y}\\
			0 & -\delta_{\alpha,z} & 0 & \delta_{\alpha,x}\\
			0 & \delta_{\alpha,y} & -\delta_{\alpha,x} & 0
			\end{pmatrix}\,, \label{eq_8a}\\
			\hat{M}_\alpha = {} & \begin{pmatrix}
			0 & \delta_{\alpha,x} & \delta_{\alpha,y} & \delta_{\alpha,z}\\
			\delta_{\alpha,x} & 0 & 0 & 0\\
			\delta_{\alpha,y} & 0 & 0 & 0\\
			\delta_{\alpha,z} & 0 & 0 & 0
			\end{pmatrix}\,.\label{eq_8b}
			\end{align}
		\end{subequations} 
	
				We emphasize that the chain Hamiltonian $\hat{\mathcal{H}}^{\text{ch}}_{\text{eff}}$ induces
				only a slow dynamics because the coupling between the head of the chain and its next site
				is $J_{\mathrm{Q}}=1$, while the coupling between different chain sites as well as the hopping processes between 	different flavors at each site is of order $\sqrt{\gamma}J_{\mathrm{Q}} \thickapprox 10^{-2}$ to $10^{-3}$. Therefore, the quantum effects such as the dynamics in the bath and eventually dephasing and relaxation of the polarization of the central spin due to the presence of the bath of spins is slow. 
				
Finally, we state that the autocorrelation expressed by the derived effective model reads 
\begin{equation}
\label{eq:auto_s_eff1}				
S^\alpha(t) = \frac{1}{4} \langle  e_\alpha,\textbf{0} | e^{-{\tt i}\hat{\mathcal{H}}_{\rm eff}}  |e_\alpha,\textbf{0}\rangle\, ,			
\end{equation}
with $e_\alpha=(0,\delta_{\alpha,x},\delta_{\alpha,y},\delta_{\alpha,z})^\top$ and $\textbf{0}$ being the vacuum
of all bosons. The autocorrelation \eqref{eq:auto_s_eff1} can be reformulated with the help of the matrix 
$\hat M_\alpha$ and $e_1=(1,0,0,0)^\top$
\begin{subequations}
\label{eq:auto_s_eff2}	
\begin{align}	
S^\alpha(t) &= \frac{1}{4} \langle  e_1,\textbf{0} | \hat M_\alpha e^{-{\tt i}\hat{\mathcal{H}}_{\rm eff}} 
\hat M_\alpha|e_1,\textbf{0}\rangle		\, ,
\\
&= 	\frac{1}{4} \langle  e_1,\textbf{0} |e^{{\tt i}\hat{\mathcal{H}}_{\rm eff}}  \hat M_\alpha
e^{-{\tt i}\hat{\mathcal{H}}_{\rm eff}} \hat M_\alpha|e_1,\textbf{0}\rangle	\, ,
\\
&= 		\frac{1}{4} \langle  e_1,\textbf{0} | \hat M_\alpha(t) \hat M_\alpha(0)|e_1,\textbf{0}\rangle	\, ,
\label{eq:auto_s_eff2_c}
\end{align}
\end{subequations}
where we used the fact that $\hat{\mathcal{H}}_{\rm eff}|e_1,\textbf{0}\rangle = 0$ since $\hat K_\alpha e_1=0$
and all bosonic terms in the chain part annihilate the bosonic vacua.

		\subsection{The bosonic truncated Wigner approximation
		\label{s3c}}
		
		In order to apply a TWA to the effective model defined in the previous
		section we need to represent the four-dimensional impurity by objects which have classical counterparts.
		Here we choose two spins with $S=1/2$ which together span a four dimensional Hilbert space. We
		denote their singlet state by $\ket{s}$ and their three triplet states by $\ket{t_\alpha}$ for 
		$\alpha\in\{x,y,z\}$, identified with the four-dimensional Cartesian vectors 
		$\ket{s} = \begin{pmatrix} 1 & 0 & 0 & 0\end{pmatrix}^\top$ and 
		$\ket{t_\alpha} = \begin{pmatrix} 0 & \delta_{\alpha x} & \delta_{\alpha y} & \delta_{\alpha z}\end{pmatrix}^\top$.
		Elementary linear algebra~\cite{PhysRevB.41.9323} yields the action of the spin operators 
		on these states
			\begin{subequations} 
			\label{eq_9}
			\begin{align}
			\hat{S}_{\nu,\alpha} \ket{s}= {} & -\frac{(-1)^{\nu}}{2} \sum_{\beta} \delta_{\alpha \beta} \ket{t_\beta}\, ,
			\label{eq_9a}\\
			\hat{S}_{\nu,\alpha} \ket{t_\beta}= {} & -\frac{1}{2} \big[2 (-1)^{\nu} \delta_{\alpha \beta} \ket{s} 
			- {\tt i}\sum_{\delta} \epsilon_{\alpha \beta \delta} \ket{t_\delta}\big]\, ,
			\label{eq_9b}
			\end{align}
		\end{subequations}
		where $\nu = \{1,2\}$ labels the spin $\hat{S}_1$ and $\hat{S}_2$, respectively.	With these definitions,
		 the matrices $\hat{K}$ and $\hat{M}$ in Eqs.~\eqref{eq_8a} and~\eqref{eq_8b} can be expressed 
		in terms of these spin operators
		\begin{subequations} 
		\label{eq_10}
			\begin{align}
			\hat{K}_\alpha = {} & - (\hat{S}_{1,\alpha} + \hat{S}_{2,\alpha})\, ,\label{eq_10a}\\
			\hat{M}_\alpha = {} & \hat{S}_{1,\alpha} - \hat{S}_{2,\alpha}\, .\label{eq_10b}
			\end{align}
		\end{subequations}
		
		The annihilation and creation operators of the harmonic oscillators can be expressed by
		position and momentum operators in the standard way
			\begin{subequations} \label{eq_11}
			\begin{align}
			\hat{r}_{j,\alpha}=\frac{1}{\sqrt{2}}(\hat{a}^{\dagger}_{j,\alpha}+\hat{a}_{j,\alpha})\, ,\label{eq_11a}\\
			\hat{p}_{j,\alpha}=\frac{{\tt i}}{\sqrt{2}}(\hat{a}^{\dagger}_{j,\alpha}-\hat{a}_{j,\alpha})\, .\label{eq_11b}
			\end{align}
		\end{subequations}
		With these relations, the Hamiltonian in Eq.~\eqref{eq_6} can be rewritten into
		\begin{subequations}\label{eq_12}
			\begin{align}
			\hat{\mathcal{H}}^{\text{CS}}_{\text{eff}}=& {}- \frac{1}{\sqrt{2}} (\hat{\vec{S}}_1 + \hat{\vec{S}}_2)\cdot \hat{\vec{r}}_1\, ,\label{eq_12a}\\
			\hat{\mathcal{H}}^{\text{ch}}_{\text{eff}}=& \frac{1}{2} \sum^{N_{}\rm tr}_{j = 1} 
			\hspace{-0.05cm}(\hat{\vec{S}}_2 - \hat{\vec{S}}_1) \hspace{-0.05cm}\cdot \hspace{-0.05cm} 
			[(\chi_j \hat{\vec{r}}_j \hspace{-0.05cm} + \eta_{j-1} \hat{\vec{r}}_{j-1} \hspace{-0.05cm}
			+ \eta_{j} \hat{\vec{r}}_{j+1})\hspace{-0.05cm}\times \hat{\vec{p}}_j],\label{eq_12b}\\
			\hat{\mathcal{H}}^{\text{Z}}_{\text{eff}} = & {} h (\hat{\vec{S}}_{1,z} + \hat{\vec{S}}_{2,z})\, .
			\label{eq_12c}
			\end{align}
		\end{subequations}
				The ensuing time evolution of the operators $\hat{\vec{r}}$, $\hat{\vec{p}}$, $\hat{\vec{S}}_1$, and $\hat{\vec{S}}_2$ according to the Heisenberg equation of motion reads 
		\begin{subequations}\label{eq_13}
			\begin{align}
			\frac{\rm d}{{\rm d} t} \hat{\vec{r}}_1 = {} & \frac{\chi_1}{2} (\hat{\vec{S}}_2 - \hat{\vec{S}}_1) 
			\times \hat{\vec{r}}_1 + \frac{\eta_1}{2} (\hat{\vec{S}}_2 - \hat{\vec{S}}_1) \times \hat{\vec{r}}_2 \, ,
			\label{eq_13a}\\
			\frac{\rm d}{{\rm d} t} \hat{\vec{p}}_1 = {} & \frac{\chi_1}{2} (\hat{\vec{S}}_2 - \hat{\vec{S}}_1) 
			\times \hat{\vec{p}}_1 + \frac{\eta_1}{2} (\hat{\vec{S}}_2 - \hat{\vec{S}}_1) \times \hat{\vec{p}}_2 
			\notag \\ {} &+ \frac{1}{\sqrt{2}} (\hat{\vec{S}}_2 + \hat{\vec{S}}_1)\, ,
			\label{eq_13b}
			\end{align}
		\end{subequations}
		for $j=1$ while for general $j>1$ we obtain	
			\begin{subequations}\label{eq_14}
			\begin{align}
			\frac{\rm d}{{\rm d} t} \hat{\vec{r}}_j = {} & \frac{\chi_j}{2} (\hat{\vec{S}}_2 - \hat{\vec{S}}_1) 
			\times \hat{\vec{r}}_j + \frac{\eta_j}{2} (\hat{\vec{S}}_2 - \hat{\vec{S}}_1) \times \hat{\vec{r}}_{j+1} 
			\notag \\ {} &+ \frac{\eta_{j-1}}{2} (\hat{\vec{S}}_2 - \hat{\vec{S}}_1) \times \hat{\vec{r}}_{j-1} \, ,
			\label{eq_14a}\\
			\frac{\rm d}{{\rm d} t} \hat{\vec{p}}_j = {} & \frac{\chi_j}{2} (\hat{\vec{S}}_2 - \hat{\vec{S}}_1) 
			\times \hat{\vec{p}}_j + \frac{\eta_j}{2} (\hat{\vec{S}}_2 - \hat{\vec{S}}_1) \times \hat{\vec{p}}_{j+1} 
			\notag \\ {} &+ \frac{\eta_{j-1}}{2} (\hat{\vec{S}}_2 - \hat{\vec{S}}_1) \times \hat{\vec{p}}_{j-1}\, ,
			\label{eq_14b}\\
			\frac{\rm d}{{\rm d} t} \hat{\vec{S}}_\nu = {} & \frac{1}{\sqrt{2}} \hat{\vec{S}}_\nu 
			\times \hat{\vec{r}}_1 + \frac{3-2\nu}{2} \hat{\vec{S}}_\nu \times \sum^{N_{\rm tr}}_{j = 1} 
			\big[ \chi_j  (\hat{\vec{r}}_j \times \hat{\vec{p}}_j) \notag \\ {} & + \eta_j  (\hat{\vec{r}}_{j+1} 
			\times \hat{\vec{p}}_j) + \eta_{j-1}(\hat{\vec{r}}_{j-1} \times \hat{\vec{p}}_j) \big] 
			- h \hat{\vec{S}}_\nu \times {\vec{z}} \, ,\label{eq_14c}
			\end{align}
		\end{subequations}
		where we use ${\vec{z}} = \begin{pmatrix}  0 & 0 & 1	\end{pmatrix}^\top$ in the last term of 
		Eq.~\eqref{eq_14c}. The sought autocorrelation \eqref{eq_3} has been  expressed 
		for the effective model in \eqref{eq:auto_s_eff2_c} which implies 
				\begin{equation}\label{eq_15}
		S^\alpha(t) = {}\frac{1}{4}\langle 		(\hat{S}^\alpha_{1}(t) - \hat{S}^\alpha_{2}(t))(\hat{S}^\alpha_{1}(0) 
		- \hat{S}^\alpha_{2}(0))\rangle\, ,
		\end{equation}
		where the expectation value is taken with respect to the singlet state of spin 1 and 2 and the
		bosonic vacua.
		
		Applying the standard TWA~\cite{POLKOVNIKOV20101790}, the
		leading quantum corrections are recovered by averaging classical trajectories over distributions of initial 
		conditions. The equations of motions \eqref{eq_13} and \eqref{eq_14} are viewed as differential equations
		for classical vectors starting from random initial conditions.
		For this purpose, normal distributions have turned out to be particularly suitable for the initial conditions.
		Their asset is that only the mean value and the variance are needed to determine the distribution fully.
		We choose a normal distribution for spin $\vec S_1$ with vanishing mean value and variance $1/4$ for each 
		component because $(\hat S^\alpha)^2=1/4$ for $S=1/2$~\cite{PhysRevLett.114.045701}. 
				Since we mimic  a singlet state $\vec S_2$ is
		always chosen to be $-\vec S_1$ initially. 
		
		The position and momentum components are also drawn from a normal distribution with vanishing means. The variances
		are straightforwardly computed considering \eqref{eq_11} yielding 
		$\langle \hat r_{j,\alpha}^2\rangle=1/2 = \langle \hat p_{j,\alpha}^2\rangle$.
		In practice, the time-evolution of the central spin $S^\alpha(t)$ in Eq.~\eqref{eq_15} is calculated for
		configuration average over $\mathcal{M}$ classical trajectories with  $\mathcal{M}$ being of the order of 
		$10^6 - 10^7$ to keep statistical errors low.

		
		\section{Numerical results}
		\label{s:results}
		
		Here we show results of the two TWAs which are the protagonists of this study. The sTWA relies
		on the classical equations of motion for the spin operators of original CSM. Either each spin
		is tracked individually or a hierarchical chain representation is used. This does not make any discernible
		difference. In contrast,  the bTWA 
		solves the classical equations of motion for the effective model obtained by mapping
		the large spin bath to a bath of bosons.
					
		Since $J_{\mathrm{Q}}$ is the energy unit in the numerical calculations, all times are measured in units of $
		1/J_{\mathrm{Q}}$ having set $\hbar$ to unity. The equations of motion have no lower or upper validity 
		cutoff in time and, thus, can be applied to discuss the spin-spin correlation from $t = 0$ to $ t \to \infty$.
		The effective number of coupled spins $N_{\rm eff}$ can also be chosen arbitrarily, but we keep in mind
		that the mapping to the effective model becomes exact in the limit of large spin baths. Further details of the effect of $N_{\rm eff} = 2/\gamma$ in the bTWA are provided in App.~\ref{ap1}.
		
	 Figure~\ref{fig2} shows the autocorrelation of the central spin 
	in absence of external magnetic fields. This is the
	central result of this paper. Clearly, we see that both approaches, sTWA and bTWA, are converged with respect
	to the truncation level $j_{\rm max}$~(for further details of the effect of $j_{\rm max}$ in the bTWA, see App.~\ref{ap2}). The curves for $j_{\rm max}=16$ do not differ discernibly from those
	for $j_{\rm max}=32$. In the inset, we focus on the behavior on short to moderate times. Here the agreement between
	both approaches is very good. Since we know from previous studies \cite{PhysRevB.90.064301} that the sTWA
	represents the quantum mechanical result very well we deduce that the bTWA also works well in this temporal regime.
\begin{figure}[b]
	\centering
	\includegraphics[width=1\linewidth]{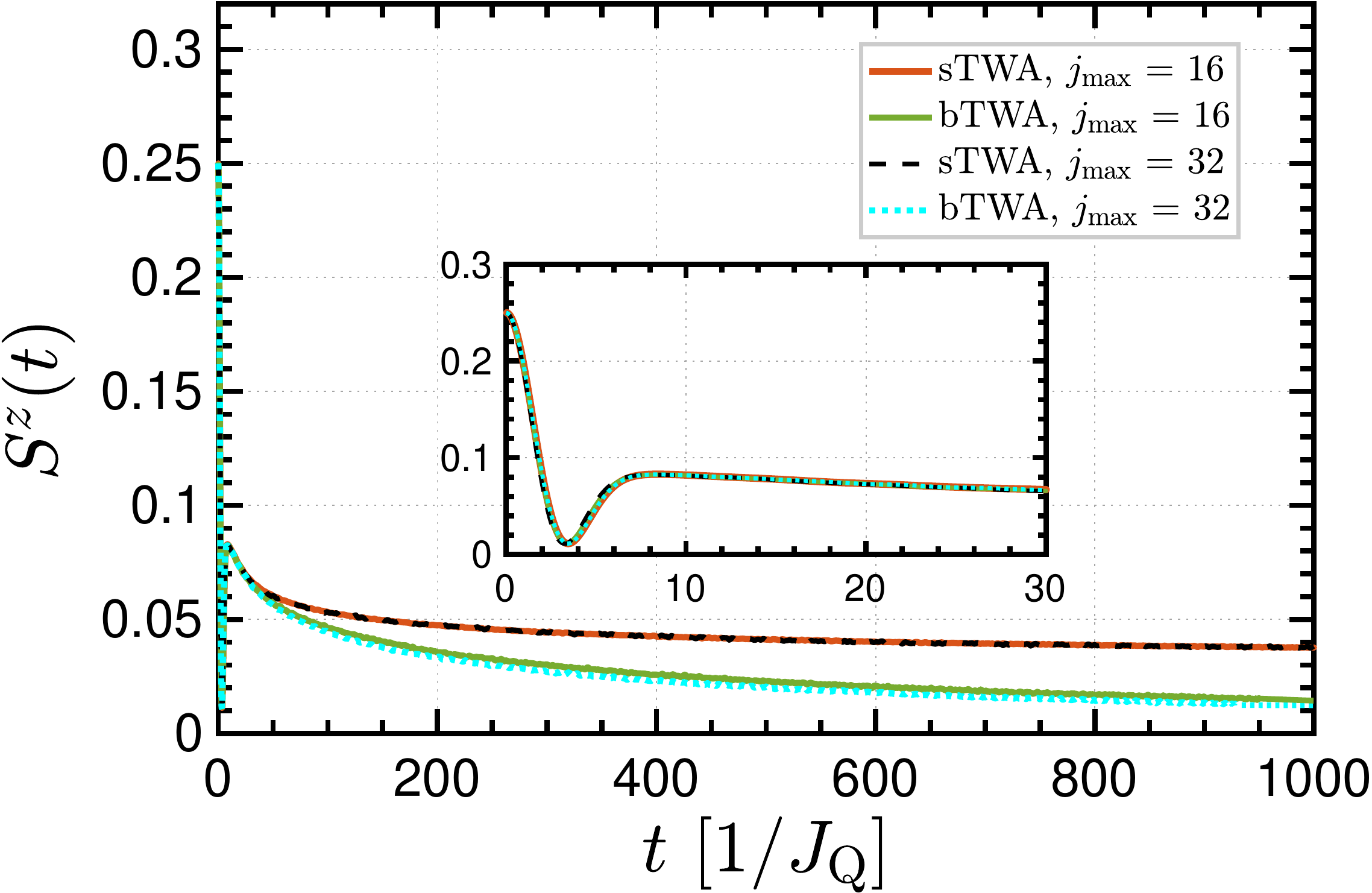}
	\caption{
		Comparison of $S^z(t)$ obtained by sTWA and by bTWA for the truncation levels 
		$j_{\rm max} = 16$ and 32, fixed number of bath spins $N = 1000$, $\gamma = 0.01$~($N_{\rm eff} = 200$), and 
		zero external magnetic field.}
	\label{fig2} 
\end{figure}

In the main panel of Fig.~\ref{fig2} we discern a significant discrepancy between the sTWA and the bTWA. This is quite surprising in view of the nice agreement up to $t\approx30/J_{\rm Q}$. The
convincing results obtained previously with sTWA~\cite{PhysRevB.90.064301} 
agrees with rigorous bounds \cite{uhrig14a,PhysRevB.94.094308} indicating a very slow decay
of the autocorrelation. Thus, the conclusion is indicated that the bTWA does not approximate the long-time
behavior of the CSM well. Still, it is (i) desirable to corroborate this conclusion further
and (ii) important to understand whether the mapping to the effective bosonic model 
introduces the observed difference or whether it is the TWA applied to the bosonic model
which induces this discrepancy. 
	
		\begin{figure}[t]
			\centering
			\includegraphics[width=1\linewidth]{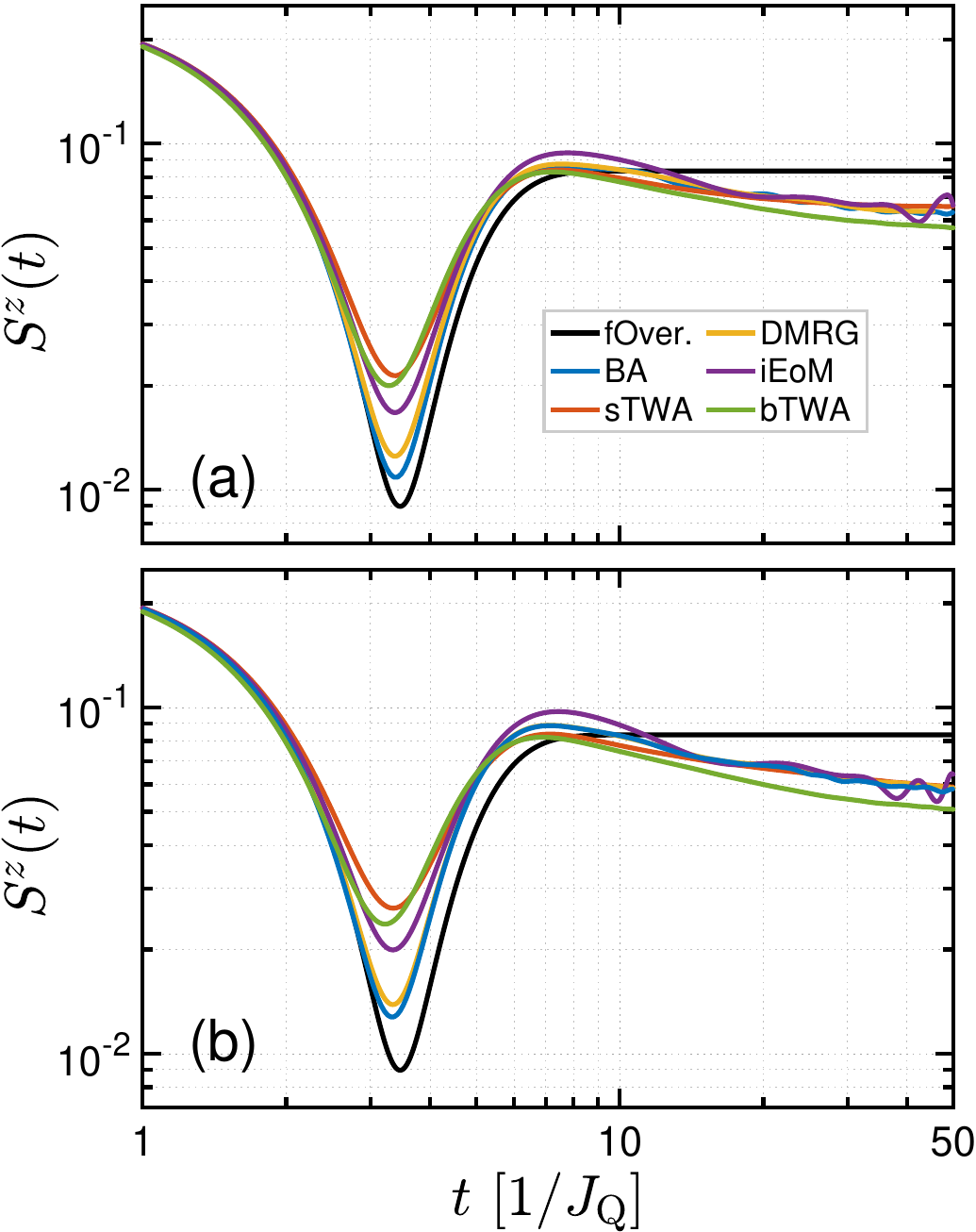}
			\caption{
			Comparison of $S^z(t)$ from various approaches~ (BA, sTWA, DMRG, and iEoM) for fixed number of bath
			spins $N = 36$ and two different (a) $\gamma = 1/18$~($N_{\rm eff} = N = 36$) and (b) 
			$\gamma = 1/12$~($N_{\rm eff} = 24$), see App.~\ref{ap1} for further details of the effect
			of $\gamma$ in the bTWA. The analytic data for random static (frozen) 
			Overhauser field (fOver) from Eq.~\eqref{eq:frozen_overhauser} 
			is included for comparison as well. In both iEoM and the TWAs, we use the truncation 
			level $j_{\rm max} = 3$, see App.~\ref{ap2} for further details of the effect of $j_{\rm max}$ in the bTWA.}
			\label{fig3} 
		\end{figure}	
	
Among the other approaches we employ the Bethe ansatz~(BA) from which we use the data published in
Ref.~\onlinecite{PhysRevB.94.094308}. The BA works perfectly for long times, but only for a moderate number of 
bath spins. Second, in  systematically controlled numerical DMRG
calculations  we consider 4096 states~\cite{PhysRevB.88.155305} with a threshold of 0.001 for the accumulated discarded weight with second-order Trotter-Suzuki decomposition. The DMRG is not able to follow the dynamics for long
 times due to the rapid growth of entanglement. But up to $t\approx 50/J_{\rm Q}$ it is reliable. 
The quantum mechanical evaluation of the iEoM  up to $j_{\rm max}=3$  with \{181,8,1\} number of bosons, respectively,
yields reliable data as well up to $t\approx 50/J_{\rm Q}$~\cite{PhysRevB.97.165431}.
Data from these methods are depicted in Fig.~\ref{fig3} for two different sets of $N$ and $N_{\rm eff}$.
The results from BA and DMRG agree very nicely for all times except for a tiny discrepancy
at the minimum which we attribute to numerical inaccuracies. Note that the BA is evaluated
based on Monte Carlo importance sampling implying small statistical fluctuations~\cite{farib13a,farib13b}.

 The iEoM approach, i.e., the 
quantum mechanical evaluation of the effective bosonic model also agrees well with the BA and DMRG data,
in particular for the slow decay beyond $t\approx 6/J_{\rm Q}$. Only the wiggles at $t\approx 50/J_{\rm Q}$
indicate that the evaluation with the limited number of bosons is at the verge of its validitiy
at this time. The discrepancies of the iEoM data to BA and DMRG
data can be attributed to the fact that the mapping to the effective model is valid for large spin baths
only, see the discussion in Ref.~\onlinecite{PhysRevB.97.165431}.
The sTWA data does not capture the minimum particularly well, but it agrees with the other approaches (BA, DMRG, iEoM)
for longer times. The frozen Overhauser data from Eq.~\eqref{eq:frozen_overhauser} 
is characterized by the constant plateau for long times because no dynamics of the Overhauser field is included.

What is the behavior of the data from bTWA? As we have already seen in Fig.~\ref{fig2} for short and moderate
times the agreement with sTWA and thus with the other data is good. In view of the long-time discrepancy
observed in Fig.~\ref{fig2} we focus on the longer times beyond $20/J_{\rm Q}$. We discern that the
data from bTWA clearly lies below the other data which coincide very well (except for the frozen
Overhauser curve). This observation corroborates our finding in Fig.~\ref{fig2} that the TWA
applied to the effective bosonic model does not approximate the long-time behavior reliably.
In addition, we learn that the iEoM data, i.e., the quantum mechanical evaluation of the effective bosonic
model, works fine at these times. Hence, Fig.~\ref{fig3} provides evidence that it is not the
mapping to the effective bosonic bath which is responsible for the discrepancy, but the bTWA.
Hence, the two questions posed above are answered. 
\begin{figure}[t]
	\centering
	\includegraphics[width=1\linewidth]{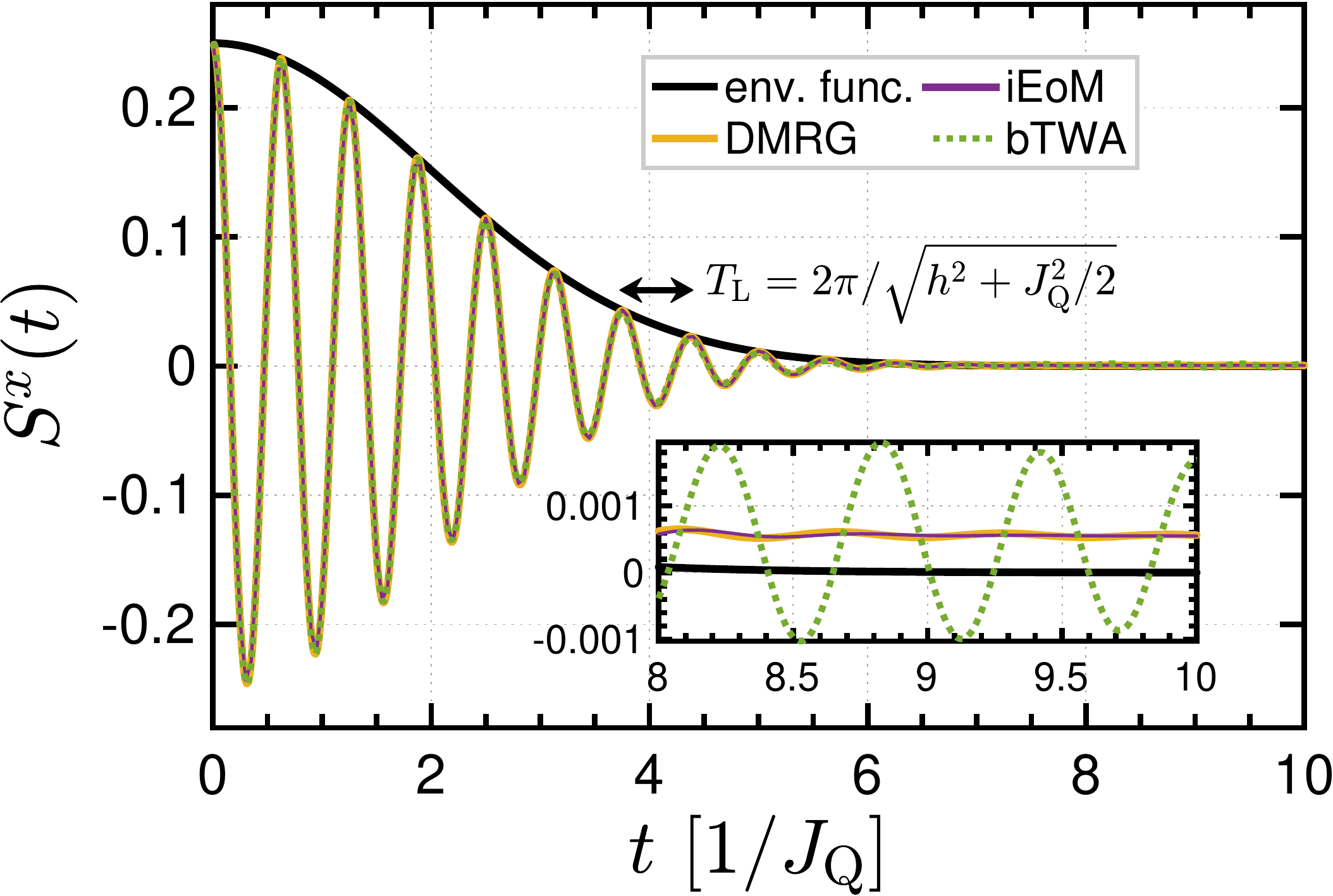}
	\caption{Comparison of the $x$-component of the spin-spin autocorrelation obtained from DMRG, iEoM, and 
		the bTWA at finite magnetic field $h/J_{\mathrm Q} = 10$ along the $z$-direction. The 
		parameters are $N= 500$, $j_{\rm max} = 3$ and $N_{\rm eff} = 200$. The period of the Larmor precession 
		is given by $T_{\rm L} = 2 \pi/\sqrt{h^2 + J^2_{\mathrm Q}/2} \approx 0.63/ J_{\mathrm Q}$. 
		The envelope function shown as black line is given by 
		$S_{\rm env.\,\,func.}(t) = \frac{1}{4}\, \exp(-J^2_{\mathrm Q} t^2/8)$. For the effect of $h$ as well as 
		the $z$-component of the spin-spin autocorrelation obtained from the bTWA, see App.~\ref{ap3}.}
	\label{fig4} 
\end{figure}

This raises the question why the TWA is not as efficient
as it is when applied directly to the spins. We do not yet have a concluding answer but the hypothesis suggesting itself is that the conserved quantities of 
the quantum effective bosonic model and its classical counterparts are not the same.
In the CSM, the conserved quantities of the quantum and of the classical model
are the same which makes their dynamics very similar \cite{PhysRevB.90.064301}.

Finally, we address the CSM in a finite magnetic field which has been well investigated both theoretically and experimentally~\cite{PhysRevB.94.245308,PhysRevB.96.115303}. Data from DMRG, iEoM, and bTWA is depicted in Fig.~\ref{fig4} for a magnetic field in $z$-direction. 
In the main panel, all data sets agree very well. 
All of them show the clear signature of Larmor precession with a 
period $T_{\rm L} = 2 \pi/\sqrt{h^2 + J^2_{\mathrm Q}/2} \approx 0.63/ J_{\mathrm Q}$, cf.~
Refs.~\onlinecite{PhysRevB.89.045317, PhysRevB.90.064301}.
The envelope function of the Larmor precession is given by 
$S_{\rm env.\,\,func.}(t) = \frac{1}{4}\, \exp(-J^2_{\mathrm Q} t^2/8)$~\cite{PhysRevB.65.205309}.

If we zoom far into the behavior at longer times after the signal has dephased, only
minor discrepancies occur. This behavior is not unexpected since we learned already in the previous
figures that the bTWA works well for times below $30/J_{\rm Q}$. Hence the Larmor precessions
and the Gaussian dephasing as shown by the black envelope function are retrieved reliably.
Only the small discrepancies at later times indicate that the approximate treatment is
not perfect at long times. But in a magnetic field the signal has essentially vanished
anyway in the long-time regime.

		\section{Summary and discussion}
		\label{s:summary}
		
		In this article, we theoretically studied the spin dynamics of the central spin in the central
		spin model (CSM). The CSM describes a so-called central spin coupled to 
		spins in its environment in a star-like topology, i.e., without coupling between pairs of bath spins.
		This model is relevant for a plethora of physical systems where a small quantum system is coupled
		to a bath of other small quantum systems. A particularly interesting framework is the realization
		of quantum bits and their decoherence mechanisms due to their interaction with spin baths.
		
		For many phenomena the long-time dynamics of large spin baths needs to be described reliably
		which poses an insurmountable challenge to brute force numerical approaches because of the
		exponential growth of the quantum Hilbert space. Hence, accurate, systematically controlled
		approximative approaches are needed. One of them is the mapping of the CSM with a large spin
		bath to a bath of bosons, i.e., to an effective bosonic model, including a four-dimensional 
		impurity at the head of the chain. The bosonic degrees of freedom can be represented in a star topology
		or in a chain topology~\cite{PhysRevB.97.165431}. The latter has the advantage that one
		can add site by site of the chain in order to reach a reliable description up to longer and longer
		times. Thus, we employed this representation here. Still, the quantum mechanical
		evaluation of the resulting central spin dynamics is a great numerical challenge. For this reason,
		we studied in the present article how well a truncated Wigner approximation~(TWA) for the bosonic
		effective model, dubbed bTWA, captures the sought dynamics. This kind of approximation 
		averages correlations of classical trajectories over distributions of initial conditions
		and describes leading quantum correlations in this way~\cite{POLKOVNIKOV20101790}.
		
		We found that the bTWA works very nicely for short and moderate times if the spin bath is
		large. This  condition on the size of the spin bath does not result from the TWA, 
		but from the mapping of the CSM to
		the effective bosonic model. Only a few bosonic sites in the chain representation of the
		bosonic bath are necessary.
		
		Much to our surprise, however, we found a qualitative discrepancy of the bTWA results compared to
		other approaches at long times. In this regime, the bTWA results display a significantly faster
		decay than the results by a direct application of the TWA to the CSM, dubbed sTWA. This discrepancy
		does not stem from the sTWA, but from the bTWA. Inspecting and comparing the behavior at moderate
		times where results from other approaches such as Bethe ansatz and DMRG are available indicates
		clearly that the correlations from bTWA are the deviating ones which are decaying too fast. 
		Although the origin of this unexpected discrepancy is still unclear, we presume that the \emph{classical} effective bosonic
		model, from which the trajectories are derived, that are averaged in bTWA over initial conditions,
		has different, probably less, conserved quantities than the quantum effective bosonic model or the original CSM.
		Note that the quantum and the classical CSM share the same conserved 
		quantities~\cite{PhysRevB.90.064301,uhrig14a,PhysRevB.94.094308}
		so that their very similar behavior is plausible.
		
		But clearly, further studies are called for to (i) identify unambiguously the origin of
		the discrepancy and (ii) to conceive reliable and efficient evaluation techniques for the
		effective bosonic model. One idea suggesting itself is to use numerical renormalization group
		techniques to evaluate its dynamics. Surely, this will enhance our understanding
		of decoherence and relaxation of small quantum systems suitable for realizing quantum bits
		or quantum sensors.
		
		\begin{acknowledgments}
		We would like to thank P Schering for useful discussions and for
		providing data of other approximate and exact approaches. 
		This study has been supported by the Deutsche Forschungsgemeinschaft (DFG, German Research Foundation) and 
		the Russian Foundation for Basic Research in the International Collaborative Research 
		Centre TRR 160 (GSU), by the DFG in 
		project UH 90/14-1 (TG and GSU), and by the Studienstiftung des Deutschen Volkes (KB). 
		In addition, we also thank for the computing time provided on
		the Linux HPC cluster LiDO3 at TU Dortmund University. M.Y. greatly acknowledges the financial support by the National Science Foundation through award numbers DMR-1945529, PHY-1607611 and NSF PHY1748958
		as well as from the Welch Foundation through award number AT-2036-20200401.
		\end{acknowledgments}
		
		\appendix
		
		\renewcommand\thefigure{\thesection.\arabic{figure}} 
		
		\section{Effect of the effective number of bath spins $N_\text{eff}$ in the bTWA}
		
		\label{ap1}
		\setcounter{figure}{0} 

		\begin{figure}[b]
			\centering
			\includegraphics[width=1\linewidth]{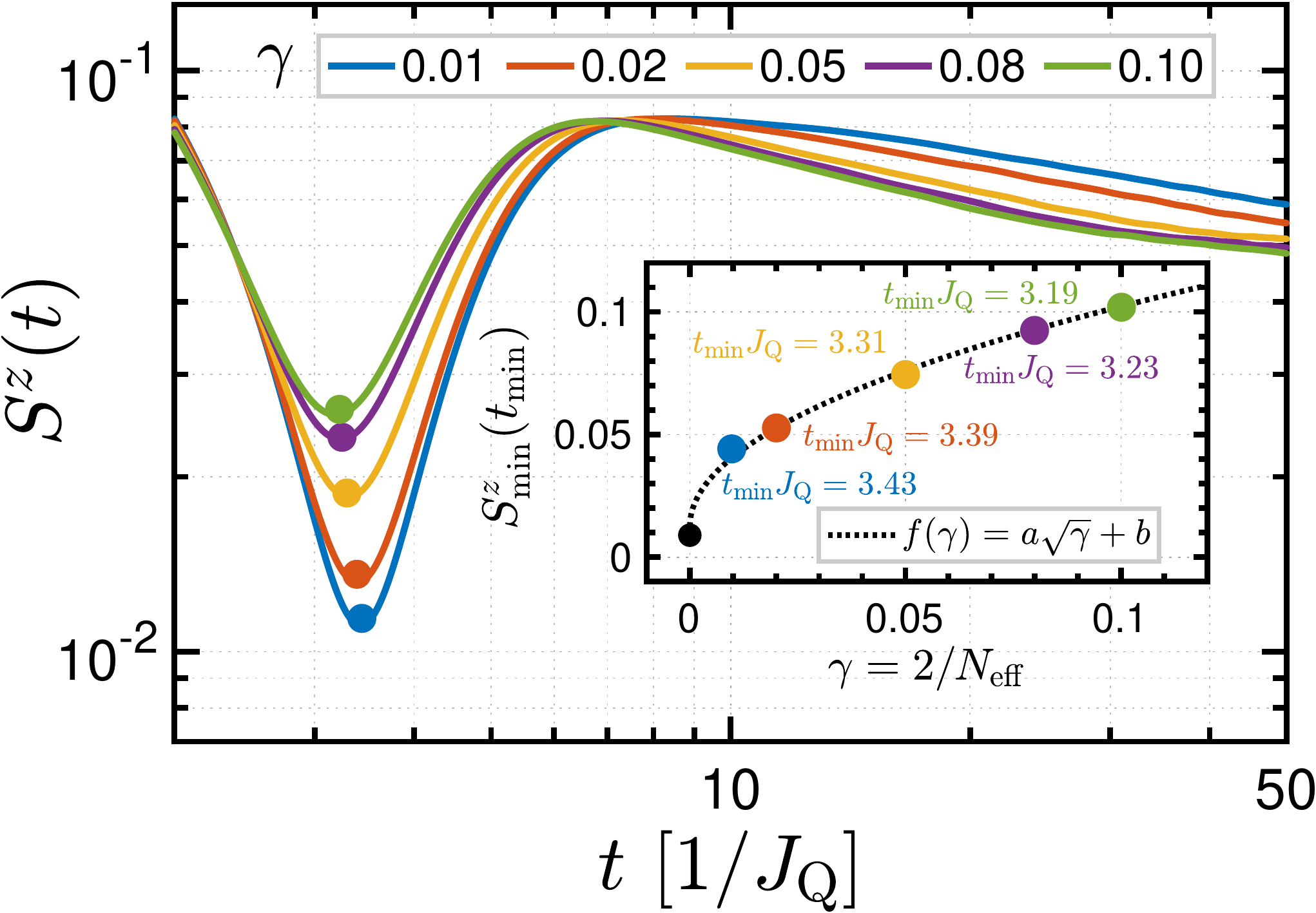}
			\caption{The effect of the effective number of bath spins characterized by 
			$\gamma = 2/N_{\rm eff}$ in the bTWA on the spin-spin correlation at fixed $j_{\rm max} = 3$ and $N = 500$. 
			The dotted {fitting function} in the inset is 
			$f(\gamma) = a \sqrt{\gamma} +b$ with $a = 0.285 \pm 0.005$ and 
			$b = S^z_{\rm min}(t_{\rm min} J_{\rm Q} = \sqrt{12})$, 
			which confirms the square root proportionality of the minimum value of the correlation on $\gamma$.}
			\label{figA} 
		\end{figure}
		
The effective number of bath spins $N_\text{eff}$ is one of the parameters influencing the minimum autocorrelation at intermediate time scales as well as the decoherence rates at long time scales. So, in the bTWA, it is important to investigate a range of $N_{\rm eff}$ for fixed $j_{\rm max} = 3$ and $N = 500$ as depicted
in Fig.~\ref{figA}, namely $N_{\rm eff} = 200$, 100, 40, 25, and 20, respectively, corresponding to 
$\gamma = 0.01$, 0.02, 0.05, 0.08, and 0.10. We obtain a square root behavior of $S^z_{\rm min}(t_{\rm min})$ as shown in the inset of Fig.~\ref{figA} for increasing $\gamma$~(decreasing $N_\text{eff}$). The coefficients $a = 0.285 \pm 0.005$ and $b = S^z_{\rm min}(t_{\rm min} J_{\rm Q} = \sqrt{12})$ in the fitting function $f(\gamma) = a \sqrt{\gamma} +b$ depend on the set of the other parameters. The spin-spin autocorrelation for $\gamma = 0$ equals the one for the frozen Overhauser field with $S^z_{\rm min}(t_{\rm min} J_{\rm Q} = \sqrt{12}) \simeq 0.009$ as a benchmark, see Eq.~\eqref{eq:frozen_overhauser}. 
This fact stems from the hyperfine coupling to the $i$-th bath that is proportional to the square root of $\gamma$. 

For larger values of $\gamma$ beyond $\simeq 0.08$ we observe that the further changes of $\gamma$ do not change the curves anymore at least up to moderate times. This observation agrees with what was found by sTWA~\cite{PhysRevB.96.054415}.

		\section{Effect of the truncation level $j_{\rm max}$ in the bTWA}
		\label{ap2}
		\setcounter{figure}{0} 
		\begin{figure}[t]
			\centering
			\includegraphics[width=1\linewidth]{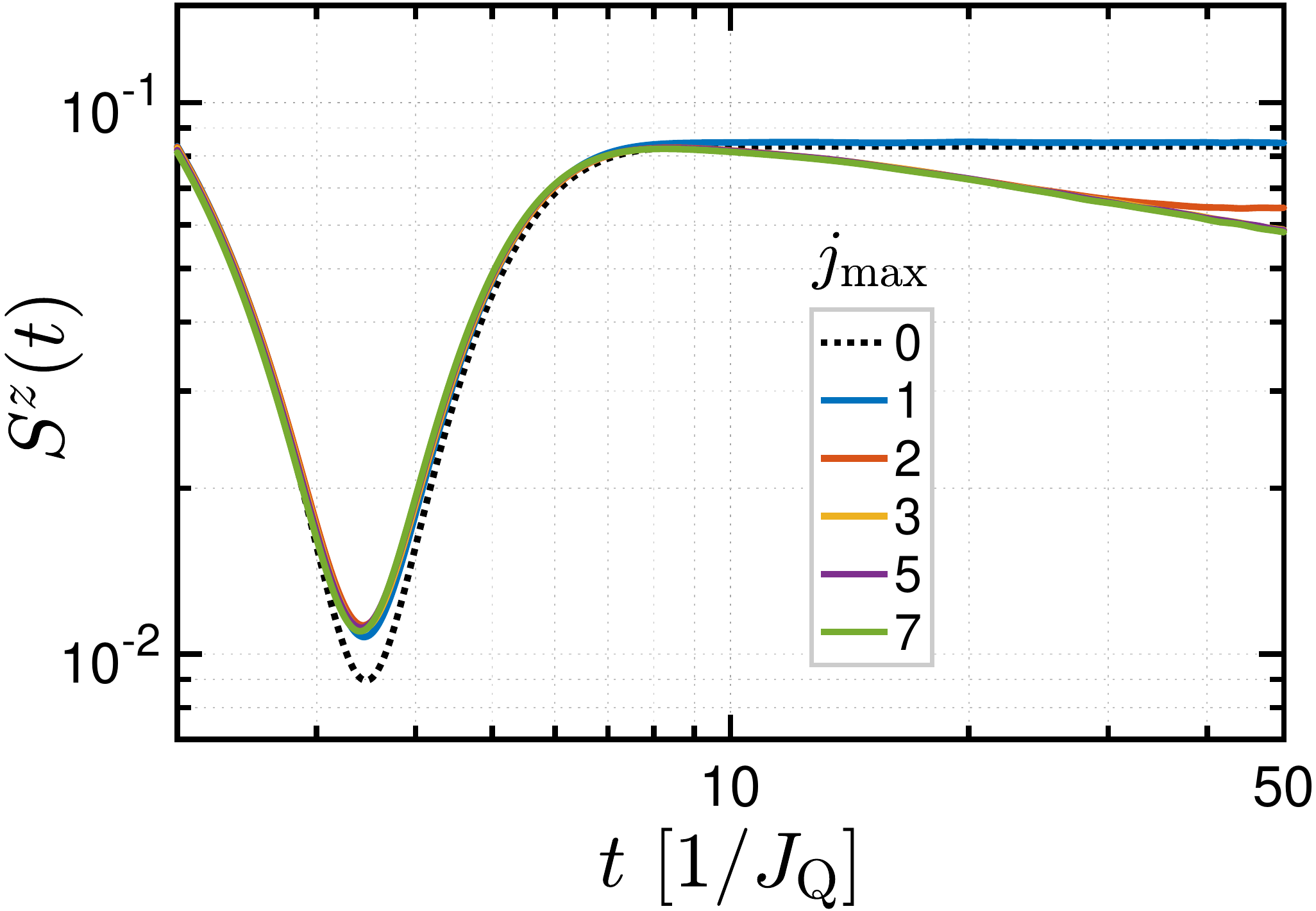}
			\caption{The effect of truncation level characterized by $j_{\rm max}$ in the bTWA 
			on the spin-spin autocorrelation at fixed $N = 500$ and $\gamma = 0.01$~($N_{\rm eff} = 200$).}
			\label{figB} 
		\end{figure}
		
		Here we study the effect of the maximum number of bosonic modes $j_{\rm max}$ in the bTWA, see
		Fig.~\ref{figB}, at fixed number of bath spins $N = 500$ and $\gamma = 0.01$~(corresponding to $N_{\rm eff} = 200$). 
		The curve for $j_{\rm max} = 0$ shows the result for the frozen Overhauser field in 
		Eq.~\eqref{eq:frozen_overhauser}. The curve for $j_{\rm max} =1$ induces only a very small temporal evolution 
		of the Overhauser bath because the central spin is  coupled only to a single harmonic oscillator which has a small
		effect on the position of the minimum. But the long-time plateau value of the autocorrelation stays close to the frozen Overhauser field one for the studied times.
		
		Taking into account a larger number of bosonic modes $j_{\rm max} \geq 2$, the difference between the static,
		frozen Overhauser result and the dynamic autocorrelations further increases. The frozen Overhauser curve~(dashed line) is always below the other curves at short timescales. Clearly, the decay of the autocorrelation sets in only for $t > \tau$ after a specific time $\tau \simeq 10/ J_{\rm Q}$ which is almost independent of the set of parameters. 
	For the shown time interval, the curves do not change significantly anymore for  $j_{\rm max} \ge 3$ in 
	accordance with previous results~\cite{PhysRevB.97.165431}.
		
		\section{Effect of the external magnetic field on the spin-spin autocorrelation in the bTWA}
		\label{ap3}
		
		In this appendix, we address the role of a longitudinal magnetic field in the bTWA with the parameters $j_{\rm max} = 3$, $N = 500$, and $\gamma = 0.01~(N_{\rm eff} = 200)$ in Fig.~\ref{figC}. In this case, the solution of Eq.~\eqref{eq_14c} displays the precession of the central spin about the effective magnetic field, i.e., the Overhauser field plus the external magnetic field. Depending on the considered spin component
the Zeeman effect implies different behavior. For the $z$-autocorrelation of the central spin, 
Fig.~\ref{figC}(a), one finds that the decoherence rate is strongly suppressed by the magnetic field in a way that it approaches zero at strong fields where the spin-spin autocorrelation becomes almost time-independent and tends to take the initial value of 1/4. This implies that the central spin polarization parallel to the external magnetic 
 field is stabilized for $h \gg J_{\rm Q}$.
					\setcounter{figure}{0} 
		\begin{figure}[t]
			\centering
			\includegraphics[width=1\linewidth]{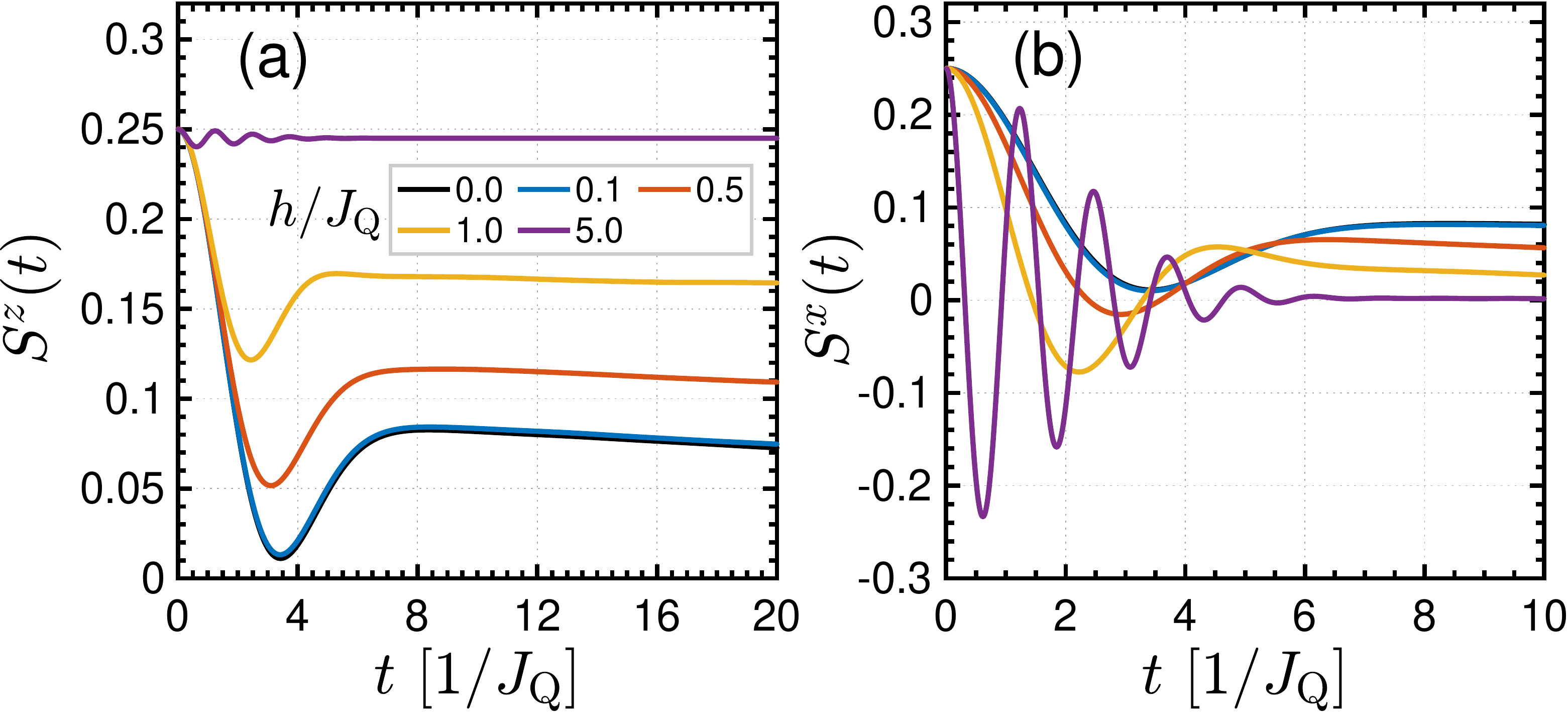}
			\caption{The effect of the external longitudinal magnetic field on the (a) 
			$z$-component and (b) $x$-component of the spin-spin autocorrelation at fixed $j_{\rm max} = 3$, $N = 500$, and 
			$\gamma = 0.01$. The longitudinal central spin polarization is stabilized with the external longitudinal field and the amplitude of the oscillations is damped for increasing $h$ such that at strong enough magnetic fields an almost  time-independent autocorrelation function 
			$S^z(t) \to 1/4$ results. In contrast, the transversal  spin-spin autocorrelation displays
			prominent Larmor precession which quickly dephase due to the random Overhauser field.}
			\label{figC} 
		\end{figure}

Upon increasing magnetic field, the minimum of the longitudinal autocorrelation occurs earlier and earlier 
before it is reduced to small oscillations and eventually to an almost constant plateau. In contrast to the longitudinal  dynamics of the central spin, the transversal dynamics, Fig.~\ref{figC}(b), displays pronounced Larmor precessions with
fast decreasing amplitude due to the dephasing induced by the fluctuations of the Overhauser field.\\
}
		
	\bibliography{bibliography.bib}
	
\end{document}